# SIMULATION OF CARDIAC FLOW: ANALYSIS OF GEOMETRY SIMPLIFICATION

Fanwei Kong (1), Christoph Augustin (2), Kevin Sack (3,4) Shawn Shadden (1)

(1) Department of Mechanical Engineering
University of California, Berkeley
Berkeley, CA, USA

(2) Institute of Biophysics
Medical University of Graz
Graz, Austria

(3) Division of Biomedical Engineering
Department of Human Biology,
University of Cape Town,
Cape Town, South Africa

(4) Department of Surgery,
University of California, San Francisco,
San Francisco, CA, United States

**INTRODUCTION**

Cardiovascular diseases (CVDs) are the leading causes of mortality worldwide. The contraction and relaxation of left ventricle (LV) is the main driving force of blood circulation. Altered LV hemodynamics is believed to be associated with the initiation and progression of many CVDs. Thus, understanding and evaluating the flow pattern inside a patient LV is thought to be essential to capture, and subsequently treat, cardiovascular dysfunction at early stages to reduce the mortality and morbidity rates [1]. Computational fluid dynamics (CFD) models, often derived from patient-specific medical imaging, have been used to provide a more fundamental understanding of individual LV flow patterns and pressure fields. Such image-based modeling may advance diagnostic capabilities, treatment protocols and help guide clinicians to choose the most effective therapy of CVDs [1].

Most prior ventricular flow studies obtained LV wall geometries from *in vivo* ultrasound-based or cardiac magnetic resonance imaging (MRI) images with limited resolution [2-3]. The model geometries were often highly simplified and usually lacked the papillary muscles (PM) and the corrugated trabecular structures of the LV. Since the LV flow pattern is sensitive to geometry, it is important to understand the effect of this simplification on modeling intraventricular flow and pressure. Here we apply CFD modeling to a subject-specific porcine LV model with detailed ventricular structures and motion obtained from previous solid mechanics finite-element (FE) simulations based on high-resolution image data [4]. We simplified the detailed LV endocardial surfaces to remove PM and trabecular structures and built a smoothed model that resembles the resolution of *in vivo* MRI images. We then compare the simulated LV flow pattern and pressure of the simplified models to those of the complex model.

**METHODS**

Detailed model geometries were extracted from the endocardial surfaces of a subject-specific FE model constructed in a previous study [4]. Briefly, a healthy porcine heart was fixed and imaged by diffusion-tensor MRI with a high spatial resolution of 0.3 × 0.3 × 0.8 mm. The constructed model preserved detailed representations of trabecular and PM structures. The biventricular model simulated realistic dynamic beating of the heart and was validated by subject-specific and independent *in vivo* strain data obtained from echocardiography. We obtained the simulated endocardial surfaces of a total of 41 time-steps for 1 cardiac cycle. We followed conventional approaches [5-6] to avoid boundary effects on the intraventricular flow: mitral opening (MO) and aortic opening (AO) were extended from the positions of valves along the axis of the aortic outflow tract.

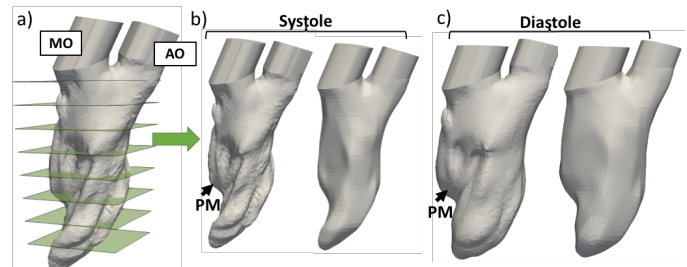

**Figure 1:** a) LV geometry at systole with cut planes, complex and smoothed geometries b) at systole and c) at diastole

To smooth out the PM and trabeculae structures, on the most contracted mesh surface, we sampled mesh points on cut planes that are 8 mm apart from each other and normal to the LV long axis (**fig. 1a**). From those sampled points, we selected those that form the vertices of the convex hull of the point set. We then down-sampled or linearly interpolated the selected points to maintain a uniform distribution of mesh vertices and applied a smoothing operation to remove sharp edges caused by mesh simplification. The same mesh vertices were chosen for other phases to maintain the mesh connectivity. **Fig. 1** display a comparison between the complex and the smoothed mesh.

We applied the Arbitrary Lagrangian-Eulerian formulation of the incompressible Navier-Stokes equations with a stabilized FE Galerkin method [7] in the open-source FEniCS project to simulate the intraventricular flow and account for the moving boundary. We created a volume mesh from one surface at the start of diastole and used cubic spline interpolation to obtain 3200 interpolated surface meshes to impose the movement on the volume mesh. **Fig. 2a** shows the LV volume curve obtained from the given boundary meshes. Diastole and systole phases were determined based on rapid increase and decrease of LV volume, while the phases when the LV volume holds steadily were determined as isovolumetric contraction (IVC), and isovolumetric relaxation (IVR) phases. Pressure boundary conditions were applied at the inlet during diastole and during IVC, and at the outlet during systole and during IVR. **Fig. 2b** displays the prescribed pressure curve on mitral inlet or aortic outlet.

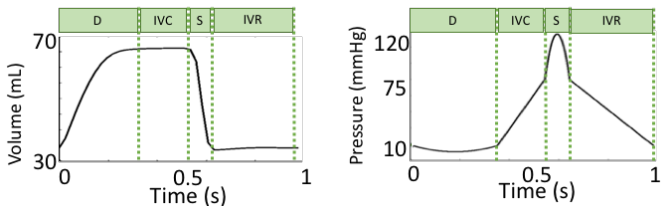

Figure 2: a) LV volume curve and b) prescribed pressure boundary condition curve of 1 cardiac cycle.

**RESULTS**

Based on the simulations, flow rates, maximum velocity at mitral or aortic openings, intraventricular flow velocity and pressure drop were extracted during the cardiac cycle (Fig. 3). The pressure drop was defined as the difference in pressure between apex and mitral opening at diastole or aortic opening at systole. The maximum velocity, flow rate and pressure drop curves displayed similar trends for both smoothed and complex models, while some differences in magnitudes were observed. During diastole and IVC, the largest discrepancy of flow rate happened after middle diastole when the flow rate for the smoothed model was 15.7% smaller. The maximum velocities at mitral opening and pressure drop were similar between the two models. At middle systole, the maximum velocity and flow rate at aortic opening and pressure drop were smaller by 5.4%, 14.3% and 26.9% respectively for the smoothed model than for the complex model. During IVR, small fluctuations of maximum velocity and pressure drop were observed in the complex model and not in the smoothed model.

The intraventricular flow patterns during diastole were different between the complex and the smoothed models (fig. 4 left). At middle diastole, for the complex model, the PM blocked the mitral jet and created disturbed low-velocity flow near the PMs. The mitral jet then formed two major vortices. One vortex located near the center of LV, sweeping the flow from the lateral wall to the septal wall. The other vortex centered at the bottom of LV, creating a circulation near the apex region. Without the PMs, intraventricular flow of the smoothed model was dominated by only one dominant vortex during diastole. However, the circulation did not seem to cover the apex region for the smoothed model. From middle to late diastole, the LV flow patterns were similar while the velocity magnitude decreased for both models. During systole (fig. 4 right), the intraventricular flow patterns were similar between the two models.

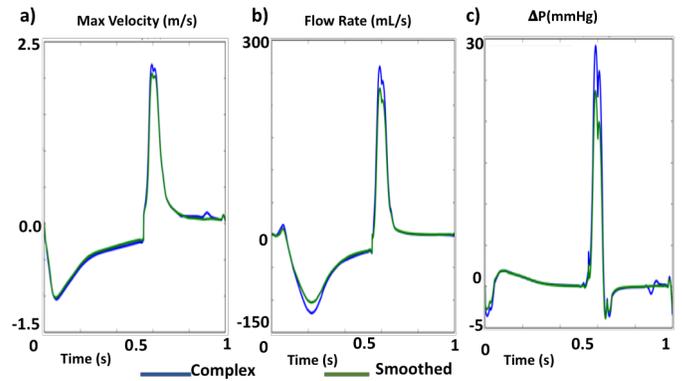

Figure 3: a) Maximum velocity and b) flow rate at mitral or aortic opening and c) pressure drop for complex and smoothed mesh

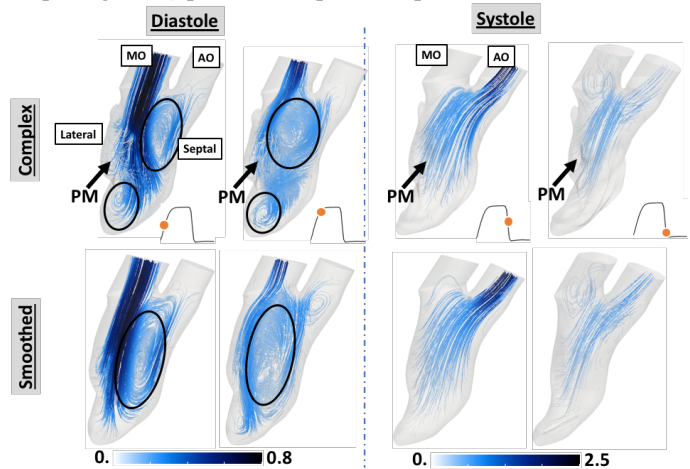

Figure 4. Comparison of velocity streamlines between complex model and smoothed model at middle diastole and middle systole. Color map represents velocity magnitude (m/s)

**DISCUSSION**

Simplified LV geometries have been commonly used in image-based CFD modeling of intraventricular blood flow. Our study provides useful insights on the effect of geometry simplification on LV CFD simulation by applying realistic LV wall motions obtained from both complex and smoothed model geometries. Major differences resulted from using the smoothed model were a) reduced pressure drop between apex and aortic opening at systole and b) the lack of complex flow pattern near PMs and the absent of circulatory flow near the apex region at diastole. These results also provide potential understanding of the functions of PM and trabeculae structures and were consistent with earlier findings that used simplified LV wall motions [5]. Future improvements would include adding left atrium, mitral valve and aorta structures for a more realistic inflow and outflow boundary condition, and prescribing physiological and subject-specific inlet or outlet pressure boundary conditions.

**ACKNOWLEDGEMENTS**

This work was supported by the National Science Foundation SI2-SSI #1663671.